\documentclass[11pt,english]{article}
\usepackage[T1]{fontenc}
\usepackage[latin9]{inputenc}
\usepackage{graphicx}
\usepackage{amssymb}

\makeatletter

\usepackage{geometry}

\usepackage{babel}
\makeatother

\begin{document}
\selectlanguage{english}

\title{NLIE of Dirichlet sine-Gordon Model for Boundary Bound States}

\author{Changrim Ahn $^{1}$, Zoltán Bajnok $^{2}$, László Palla $^{3}$,
Francesco Ravanini $^{4}$}

\maketitle
\begin{center}
\textbf{We dedicate this work to the memory of Alyosha Zamolodchikov} 
\par\end{center}

\begin{center}
\textit{$^{1}$}\emph{Department of Physics, Ewha Womans Univ., Seoul
120-750, South Korea}\\
 \textit{$^{2}$HAS Theoretical Physics Research Group,} \textit{$^{3}$Department
of Theoretical Physics, Roland Eötvös Univ. H-1117 Budapest Pázmány
s. 1/A, Hungary}\\
 \textit{$^{4}$ Univ. of Bologna, Physics Department, INFN Section,
Via Irnerio 46, Bologna 40126, Italy} 
\par\end{center}

\begin{abstract}
We investigate boundary bound states of sine-Gordon model on the finite-size
strip with Dirichlet boundary conditions. For the purpose we derive
the nonlinear integral equation (NLIE) for the boundary excited states
from the Bethe ansatz equation of the inhomogeneous XXZ spin 1/2 chain
with boundary imaginary roots discovered by Saleur and Skorik. Taking
a large volume (IR) limit we calculate boundary energies, boundary
reflection factors and boundary Lüscher corrections and compare with
the excited boundary states of the Dirichlet sine-Gordon model first
considered by Dorey and Mattsson. We also consider the short distance
limit and relate the IR scattering data with that of the UV conformal
field theory. 
\end{abstract}

\section{Introduction}

Boundary integrable field theories in two dimensions have been investigated
mainly by two approaches. The boundary bootstrap approach determines
the reflection amplitudes in a factorized $S$-matrix framework which
is valid in the large volume (IR) limit \cite{GZ}. In a small volume,
on the other hand, the underlying quantum field theories can be described
as perturbed boundary conformal field theories. Complete understanding
is possible only after the two approaches are linked in such a way
that the states and operators in the alternative formulations are
exactly matched.

The nonlinear integral equation (NLIE) has been most effective in
linking the two descriptions for the sine-Gordon model. The NLIE for
the bulk theory has been started several years ago by various authors
\cite{DdV,KBP}. An evident advantage of this method is that it can
deal with excited states relatively easily as shown with great success
in the bulk sine-Gordon model \cite{FMQR,DdV97}. The NLIE is a sort
of continuum limit of the Bethe Ansatz equation (BAE) of an inhomogeneous
alternating spin 1/2 XXZ chain model which regularizes the sine-Gordon
model while keeping integrability. This method has been extended to
the sine-Gordon model defined on a strip with two boundaries. The
ground state NLIE for the Dirichlet boundary conditions (BCs) has
been studied in \cite{LMSS} and for general non-diagonal BCs in \cite{GenNLIE}.
The bulk excited state NLIE for the Dirichlet BCs has been analyzed
in \cite{DNLIE} and the hole excited state for a general non-diagonal
BCs in \cite{GenexNLIE}. In the present paper we investigate how
this method can be extended to the boundary excited states for Dirichlet
BCs.

The complete spectrum of boundary excited states of the Dirichlet
sine-Gordon (DSG) model on a half line with one boundary has been
constructed by Dorey and Mattsson (DM) by inspecting the analytic
structure of the reflection matrix in a bootstrap approach \cite{MD}.
They found a rich structure of excited boundary states, the \emph{boundary
bound states} (BBS). These states are the scattering states which
can be defined only in the IR limit of the finite size setting. It
is important to relate the IR states with the UV conformal states
appearing in the small volume description for the complete understanding
of the DSG model. In this paper we analyze carefully the {}``imaginary
roots'' of the boundary XXZ BAE, first discovered by Saleur and Skorik
\cite{Salsko}. From this we derive the NLIE including the imaginary
roots which describes the BBSs of the DSG in the whole scale. Taking
the IR limit of the NLIE, we can show that there is a one to one correspondence
between the purely imaginary roots and the DM BBSs.

The paper is organized as follows: In section 2 we summarize the available
results of the DSG model. We start by analyzing the conditions for
the existence of the imaginary roots of the inhomogeneous spin 1/2
XXZ model. The NLIE, derived from the lattice BAE, contains parameters
originating from the XXZ model which should be compared with the bootstrap
solution \cite{MD}. We analyze the large volume limit of the NLIE
in section 3 and find agreement with the BAE classification. This
leads to full physical interpretation of the energy levels described
by the NLIE without any source and the one with imaginary roots only.
We do it in two steps, first with the simpler repulsive case and then
the more complicated attractive one. In both cases the proposed correspondence
between the various source terms and the DM spectrum of BBSs are derived
from the soliton and breather reflection amplitudes and matching of
the boundary energies. As a final check we compare the finite size
energy correction derived from the NLIE to the boundary Lüscher correction
\cite{BLusch}. Section 4 deals with the calculation of the conformal
dimensions of underlying boundary conformal field theory by taking
the UV limit of the NLIE, which gives another convincing support of
our result. We conclude and give the outlook for future investigations
in section 5.

\section{Derivation of the NLIE}

In this section we summarize the results available in the literature
tailor-made for future applications.

\subsection{Imaginary Roots of the Bethe Ansatz equation}

To derive the NLIE for the DSG model, we consider anti-ferromagnetic
XXZ spin 1/2 model in a chain of $N$ sites with lattice spacing $a$,
coupled to parallel magnetic fields $h_{-}$ and $h_{+}$ at the left
and right boundaries, respectively. Its Hamiltonian is written as
\begin{equation}
\mathcal{H}(\gamma,h_{+},h_{-})=-J\sum_{n=1}^{N-1}\left(\sigma_{n}^{x}\sigma_{n+1}^{x}+\sigma_{n}^{y}\sigma_{n+1}^{y}+\cos\gamma\sigma_{n}^{z}\sigma_{n+1}^{z}\right)+h_{+}\sigma_{1}^{z}+h_{-}\sigma_{N}^{z}.\label{eq:1}\end{equation}
 Here $\sigma_{n}^{\alpha}$, $\alpha=x,y,z$ are Pauli matrices on
the $n$-th site and the anisotropy is $0\le\gamma\pi$. Whenever
necessary, we will use another coupling constant $p$ defined by \[
p=\frac{\pi}{\gamma}-1,\quad0<p<\infty.\]

The BAEs for the boundary XXZ chain (\ref{eq:1}) have been derived
by Alcaraz et al. \cite{Alcaraz} and Sklyanin \cite{Sklyanin} using
an algebraic Bethe ansatz approach. The BAEs are coupled equations
for a set of $M$ roots which have distinct values $\theta_{1},\ldots,\theta_{M}$
with $M\le N/2$; \[
\left[s_{1}(\theta_{j}+\Lambda)s_{1}(\theta_{j}-\Lambda)\right]^{N}s_{H_{+}}(\theta_{j})s_{H_{-}}(\theta_{j})=\prod_{k=1,k\neq j}^{M}s_{2}(\theta_{j}-\theta_{k})s_{2}(\theta_{j}+\theta_{k})\]
 where we introduced a short notation \[
s_{\nu}(x)=\frac{\sinh\frac{\gamma}{\pi}\left(x+\frac{i\nu\pi}{2}\right)}{\sinh\frac{\gamma}{\pi}\left(x-\frac{i\nu\pi}{2}\right)}.\]
 The boundary parameters $H_{\pm}$ in the BAEs are related to those
in the Hamiltonian by \[
h_{\pm}=\sin\frac{\pi}{p+1}\cot\frac{2\pi(H_{\pm}+1)}{p+1}.\]
 In addition to real and complex roots, we are interested in the {}``imaginary''
roots which have vanishing real parts. These objects, first observed
by Saleur and Skorik \cite{Salsko}, depend on the boundary parameters
of both sides independently. For simplicity, we set the value of $h_{-}$
so that it does not introduce any imaginary root and recall from \cite{Salsko}
how the existence and locations of the imaginary roots depend on the
values of $h_{+}$ in the limits when $\Lambda\rightarrow\infty$
and $N\rightarrow\infty$. Defining $\kappa_{j}$ by \[
e^{\kappa_{j}}=s_{1}(\theta_{j}+\Lambda)s_{1}(\theta_{j}-\Lambda),\]
 one can see that an imaginary root with $\theta_{j}=iu_{j}$ in the
$\Lambda\rightarrow\infty$ limit satisfies \[
\kappa_{j}=2A\sin\gamma\sin\frac{2\gamma u_{j}}{\pi}+O(e^{\frac{-4\gamma\Lambda}{\pi}}),\qquad{\rm with}\qquad A=\left[\cosh\frac{2\gamma\Lambda}{\pi}\right]^{-1}.\]
 Now we look for some {}``string'' solution in the form of \[
\theta_{j}=i(-\frac{\pi H_{+}}{2}+jp\pi+\epsilon_{j}),\]
 where $\epsilon_{j}$'s are supposed to be exponentially small in
the $N\rightarrow\infty$ limit. Following \cite{Salsko}, we denote
it as {}``$(n,m)$ string'' if $j$ can take integer values from
$-n$ to $m$. Then, the BAEs take the following form \begin{eqnarray*}
 &  & e^{\kappa_{m}N}\propto\epsilon_{m}-\epsilon_{m-1}\;;\quad e^{\kappa_{m-1}N}\propto\frac{\epsilon_{m-1}-\epsilon_{m-2}}{\epsilon_{m}-\epsilon_{m-1}}\;;\;\dots\;;\; e^{\kappa_{l}N}\propto\frac{\epsilon_{l}-\epsilon_{l-1}}{\epsilon_{l+1}-\epsilon_{l}}\;;\;\dots\;;\; e^{\kappa_{0}N}\epsilon_{0}\propto\frac{\epsilon_{0}-\epsilon_{-1}}{\epsilon_{1}-\epsilon_{0}}\\
 &  & e^{\kappa_{-1}N}\propto\frac{\epsilon_{-1}-\epsilon_{-2}}{\epsilon_{0}-\epsilon_{-1}}\;;\;\dots\;;\; e^{\kappa_{-n+1}N}\propto\frac{\epsilon_{-n+1}-\epsilon_{-n}}{\epsilon_{-n+2}-\epsilon_{-n+1}}\;;\quad e^{\kappa_{-n}N}\propto\frac{1}{\epsilon_{-n+1}-\epsilon_{-n}}\end{eqnarray*}
 where we have omitted finite factors in the proportionality. Since
all the $\epsilon_{j}$ are small we can determine them recursively
starting from both $-n$ and $m$. They are consistent provided \begin{eqnarray}
\kappa_{m}<0\quad,\quad & \kappa_{m}+\kappa_{m-1}<0\quad,\dots, & \kappa_{m}+\kappa_{m-1}+\dots+\kappa_{1}<0,\label{kineqp}\\
\kappa_{-n}>0\quad,\quad & \kappa_{-n}+\kappa_{-n+1}>0\quad,\dots, & \kappa_{-n}+\kappa_{-n+1}+\dots+\kappa_{-1}>0,\label{kineqm}\end{eqnarray}
 and \[
\kappa_{-n}+\kappa_{-n+1}+\dots+\kappa_{-1}+\kappa_{0}+\kappa_{m}+\kappa_{m-1}+\dots+\kappa_{1}>0.\]
 In solving these inequalities we are interested in the domain in
$H_{+}$ for a fixed $p$ in which the $(n,m)$ string can exists.
Since at some point we want to make connection with the BBSs we introduce
the parameter: \[
H_{+}=p(1-2\xi_{+}/\pi)\]
 For small/large enough $\xi_{+}$ the boundary state is absent and
we are going to determine the value of $\xi_{+}$ at which such a
string can enter/leave the physical strip. For this we plot $\sin(\frac{2\gamma}{\pi}u_{j})$
which is the relevant part of $\kappa_{j}$ together with the $(n,m)$
string on Figure 1.

\begin{center}
\begin{figure}[bh]
\begin{centering}
\includegraphics[height=3.5cm]{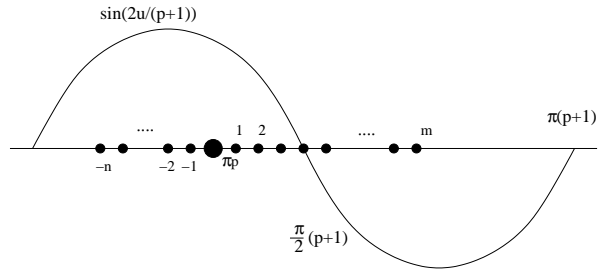}
\par\end{centering}

\caption{The boundary condition dependent factor of $\kappa_{j}$ in the $(n,m)$
string}

\end{figure}
 
\par\end{center}

\noindent From (\ref{kineqp}) we can see that such a state appears
when the average of $u_{1}$ and $u_{m}$ is exactly \[
p\xi_{0}=\frac{\pi}{2}(1+p).\]
 So for the existence we need $u_{1}+u_{m}>\pi(p+1)$ that is \[
\xi_{+}>\frac{\pi}{2p}-(m-1)\frac{\pi}{2}\]
 From (\ref{kineqm}) we can see that $u_{-n}>0$ is also needed that
is \[
\xi_{+}\geq(2n+1)\frac{\pi}{2}\]
 Finally from (\ref{kineqm}) we can see that the state disappears
when $u_{-n}+u_{m}>\pi(p+1)$ since we have too many points in the
negative part of the sine function. So for the existence of the $(n,m)$
string we also need \[
\xi_{+}<\frac{\pi}{2p}+(n+2-m)\frac{\pi}{2}.\]

\subsection{NLIE: the continuum limit of the BAE}

Based on the BAE, the NLIE equation determining the counting function
$Z(\theta)$ in the continuum limit $N\to\infty$ can be written as
\cite{DNLIE} \begin{eqnarray}
Z(\theta) & = & 2ML\sinh\theta+g(\theta|\{\theta_{k}\})+P_{{\rm bdry}}(\theta)\nonumber \\
 & - & 2i\mathrm{Im}\int dxG(\theta-x-i\epsilon)\log\left[1-(-1)^{M_{SC}}e^{iZ(x+i\epsilon)}\right],\label{NLIE}\end{eqnarray}
 where $P_{{\rm bdry}}(\theta)$ is the boundary contribution given
by \begin{eqnarray}
P_{{\rm bdry}}(\theta) & = & 2\pi\int_{0}^{\theta}dx[F(x,H_{+})+F(x,H_{-})+G(x)+J(x)]\label{Pbdry}\\
G(\theta) & = & \int_{-\infty}^{\infty}\frac{dk}{2\pi}\frac{\sinh\frac{\pi}{2}(p-1)k}{2\sinh\frac{\pi}{2}pk\cosh\frac{\pi}{2}k}e^{ik\theta},\qquad\mathrm{for}\qquad|\mathrm{Im\theta|}<\pi\min(1,p)\nonumber \\
J(\theta) & = & \int_{-\infty}^{\infty}\frac{dk}{2\pi}\frac{\sinh\frac{\pi}{4}(p-1)k\cosh\frac{\pi}{4}(p+1)k}{\sinh\frac{\pi}{2}pk\cosh\frac{\pi}{2}k}e^{ik\theta},\qquad\mathrm{for}\qquad|\mathrm{Im\theta|}<\frac{\pi}{2}\min(1,p)\nonumber \\
F(\theta,H) & = & \int_{-\infty}^{\infty}\frac{dk}{2\pi}\mathrm{sign}(H)\frac{\sinh\frac{\pi}{2}(p+1-|H|)k}{2\sinh\frac{\pi}{2}pk\cosh\frac{\pi}{2}k}e^{ik\theta},\qquad\mathrm{for}\qquad|\mathrm{Im\theta|<\frac{\pi}{2}|}H|.\nonumber \end{eqnarray}
 We have introduced a mass scale $M$ which will be identified with
that of a soliton and the finite size $L$ by \[
L=Na,\qquad M=\frac{2}{a}e^{-\Lambda}.\]
The {}``source'' term is given by \[
g(\theta|\{\theta_{k}\})=\sum_{k}c_{k}[\chi_{(k)}(\theta-\theta_{k})+\chi_{(k)}(\theta+\theta_{k})]\]
 where \[
\chi(\theta)=2\pi\int_{0}^{\theta}dxG(x)\]
 and $\{\theta_{k}\}$ is the set of position of the various objects
(holes, close and wide roots, specials) characterizing a certain state.
They satisfy the quantization rule \[
Z(\theta_{j})=2\pi I_{j}\quad,\quad I_{j}\in\mathbb{Z}+\frac{\rho}{2}\qquad\rho=M_{SC}\,\bmod\,2\quad.\]
 The coefficients $c_{k}$ are given by \[
c_{k}=\left\{ \begin{array}{ll}
+1 & \mathrm{for\, holes}\\
-1 & \mathrm{for\, all\, other\, objects}\end{array}\right.\]
 and for any function $f(\vartheta)$ we define \[
f_{(k)}(\theta)=\left\{ \begin{array}{ll}
f_{{\rm II}}(\theta) & \mathrm{for\ wide\ roots}\\
f(\theta+i\epsilon)+f(\theta-i\epsilon) & \mathrm{for\ specials}\\
f(\theta) & \mathrm{for\ all\ other\ objects}\end{array}\right.\]
 where the \emph{second determination} of $f(\theta)$ is defined
by \[
f_{{\rm II}}(\theta)=\left\{ \begin{array}{ccc}
f(\theta)+f(\theta-i\pi\ \mathrm{sign\ Im}\theta) & \mathrm{if} & p>1\\
f(\theta)-f(\theta-i\pi p\ \mathrm{sign\ Im}\theta) & \mathrm{if} & p<1\end{array}\right.\,\,{\rm for}\,|{\rm Im}\theta|>\pi\min(1,p)\quad.\]

For the vacuum state containing real roots only, Eq.(\ref{NLIE})
coincides with the one found some years ago in \cite{LMSS}. Once
the equation is solved for $Z(\theta+i\epsilon)$ one can use this
result to compute the $Z(\theta)$ function at any value in the analyticity
strip $|\mathrm{Im}\theta|<\pi\min(1,p)$, provided the function $P_{{\rm bdry}}(\theta)$
is well defined there. To extend the function outside this analyticity
strip one has to resort to the following modification of the NLIE
\begin{eqnarray*}
Z(\theta) & = & 2ML\sinh_{{\rm II}}\theta+g_{{\rm II}}(\theta|\{\theta_{k}\})+{P_{{\rm bdry}}}_{{\rm II}}(\theta)\\
 & - & 2i\mathrm{Im}\int dxG_{{\rm II}}(\theta-x-i\epsilon)\log\left[1-(-1)^{M_{SC}}e^{iZ(x+i\epsilon)}\right]\quad.\end{eqnarray*}
 The continuum limit of the counting equation which restrict the allowed
root configuration is given by \begin{equation}
N_{H}-2N_{S}=\frac{1}{2}({\rm sign}(H_{+})+{\rm sign}(H_{+}))-1+M_{C}+2M_{W}{\rm step}(p-1)+m.\label{conticount}\end{equation}
 (The integer $m$ appearing here is related to possible winding of
the sine-Gordon field, see next section).

Once $Z(\theta)$ is known, it can be used to compute the energy.
It is composed of bulk and boundary terms whose expression can be
found in \cite{LMSS} and a Casimir energy scaling function given
by \begin{equation}
E=M\sum_{k}c_{k}\cosh_{(k)}\theta_{k}-M\int\frac{dx}{4\pi}\sinh x\, Q(x).\quad;\qquad Q(x)=2\Im m\log\left[1-(-1)^{M_{SC}}e^{iZ(x+i\epsilon)}\right]\label{theenergy}\end{equation}

\subsection{Relation to Boundary sine-Gordon model}

The continuum limit of the inhomogeneous XXZ spin 1/2 chain describes
the sine-Gordon model. If we introduce diagonal ($\sigma_{z}$ only)
boundary condition on the spin chain, the continuum limit should describe
the DSG model, whose action can be written as \[
\mathcal{A}_{DSG}={\displaystyle \frac{1}{2}\
\int_{-\infty}^{\infty}dt{\displaystyle \int_{0}^{L}dx\left[\left(\partial_{\mu}\phi\right)^{2}+{\displaystyle \frac{2m_{0}^{2}}{\beta^{2}}\cos\beta\phi}\right]}}\]
 with the Dirichlet boundary conditions \[
\phi\left(0,t\right)\equiv\phi_{-}+\frac{2\pi}{\beta}m_{-}\qquad;\quad\phi(L,t)=\phi_{+}+\frac{2\pi}{\beta}m_{+}\;,\quad m_{\pm}\in\mathbb{Z}.\]
 Notice that the bulk and boundary parameters of the DSG and spin
chain models are related by \cite{LMSS} \begin{equation}
p^{-1}=\frac{8\pi}{\beta^{2}}-1,\qquad;\qquad H_{\pm}=p\left(1\mp\frac{8}{\beta}\phi_{\pm}\right).\label{boundparams}\end{equation}

This model has several important applications ranging from condensed
matter physics to string theory. An important feature of the DSG model
is the conservation of the topological charge \[
Q\equiv\frac{\beta}{2\pi}\left[\int_{0}^{L}dx\frac{\partial}{\partial x}\phi(x,t)-\phi_{+}+\phi_{-}\right]=m_{+}-m_{-}\in\mathbb{Z}\quad.\]
 The model enjoys the discrete symmetry of the field $\phi\to\phi+\frac{2\pi}{\beta}m$
and simultaneously $\phi_{\pm}\to\phi_{\pm}+\frac{2\pi}{\beta}m$
($m\in\mathbb{Z}$). The charge conjugation symmetry $\phi\to-\phi$
sending solitons into anti-solitons is also guaranteed, provided $\phi_{\pm}\to-\phi_{\pm}$
simultaneously.

The well known bulk particle spectrum of sine-Gordon model is composed
of solitons and anti-solitons with topological charge 1 and $-1$
respectively, and their bound states known as breathers in the \emph{attractive}
regime $0<\beta\leq\sqrt{4\pi}$ and of course they are also part
of the DSG spectrum. Another important part of the spectrum of the
DSG model in the half line theory - i.e. in the presence of one boundary
only - is the complicated spectrum of the BBSs described in \cite{MD}.
In addition to the bulk S-matrices \cite{zam-zam} and boundary reflection
matrices for the soliton or anti-solitons \cite{GZ}, the complete
excited boundary reflection matrices in the presence of the BBSs have
been found in \cite{MD}.

The DM result on the BBSs of the half line theory can be summarized
as follows. First, define two sets of variables \begin{eqnarray*}
\nu_{n} & = & \xi p-\frac{(2n+1)\pi p}{2}=\nu_{0}-np\pi\quad;\qquad n\geq0\\
w_{m} & = & \pi-\xi p-\frac{(2m-1)\pi p}{2}=\pi-\nu_{-m}.\end{eqnarray*}
where the bootstrap parameters $p$ and $\xi$ are related to those
of the Lagrangian as \begin{equation}
\frac{p+1}{p}=\frac{8\pi}{\beta^{2}}\qquad;\qquad\xi_{\pm}=\frac{4\pi}{\beta}\phi_{\pm}\label{eq:xiphi}\end{equation}
Whenever a condition \begin{equation}
\frac{\pi}{2}>\nu_{n_{1}}>w_{m_{2}}>\nu_{n_{2}}>\dots>w_{m_{k}}>\nu_{n_{k}}>\dots>0\label{MattsonDorey}\end{equation}
is satisfied, a BBS can exist. If the last variable is of $\nu$ type,
then the BBS is denoted as \[
|1;n_{1},m_{1},\dots,n_{k-1},m_{k-1},n_{k}\rangle\]
and if it is of $w$ type, then the state is denoted as \[
|0;n_{1},m_{1},\dots,n_{k-1},m_{k-1},n_{k},m_{k}\rangle.\]
The energy of such a state relative to the ground state is given by
\[
E_{|0/1;n_{1},m_{1},\dots,n_{k-1},m_{k-1},\dots\rangle}=\sum_{j}M\cos\nu_{n_{j}}+\sum_{j}M\cos w_{m_{j}}.\]
 The boundary reflection matrix on excited boundaries can also be
derived. We denote by $P_{|0\rangle}^{+}(\xi,\theta)$ the boundary
reflection matrix element of a soliton on the ground state boundary.
The boundary parameter dependent part reads as \cite{Salsko} \begin{equation}
-i\frac{d}{d\theta}\log\frac{P_{|0\rangle}^{+}(\theta,\xi)}{R_{0}(\theta)}=\int_{-\infty}^{\infty}dk\mathrm{e}^{ik\theta}\frac{\sinh\left(\frac{\pi}{2}(1+\frac{2\xi p}{\pi})k\right)}{2\sinh\frac{\pi}{2}pk\cosh\frac{\pi}{2}k}\label{Intreps}\end{equation}
 The reflection factor of solitons on the $|1;0\rangle$ excited boundary
is given by \begin{equation}
P_{|1;0\rangle}^{+}(\theta,\xi)=P_{|0\rangle}^{+}(\theta,\xi)a(\theta-i\nu_{0})a(\theta+i\nu_{0})\equiv P_{|1\rangle}^{+}(\theta,\xi)\label{exref}\end{equation}
 where $a(\theta)=e^{i\chi(\theta)}$ describes the soliton-soliton
S-matrix element. The reflection factor on the general boundary is
given by \[
P_{|0/1;n_{1},m_{1},\dots,n_{k-1},m_{k-1},\dots\rangle}^{+}=P_{|0/1\rangle}^{+}\prod_{k}\frac{a(\theta-i\nu_{n_{k}})a(\theta+i\nu_{n_{k}})}{a(\theta-i\nu_{0})a(\theta+i\nu_{0})}\prod_{k}\frac{a(\theta-iw_{m_{k}})a(\theta+iw_{m_{k}})}{a(\theta-iw_{0})a(\theta+iw_{0})}.\]
 Using the identity \begin{equation}
\frac{a(\theta-iw_{m_{k}})a(\theta+iw_{m_{k}})a(\theta-i(\pi-w_{m_{k}}))a(\theta+i(\pi-w_{m_{k}}))}{a(\theta-i\nu_{0})a(\theta+i\nu_{0})a(\theta-iw_{0})a(\theta+iw_{0})}=1.\label{eq:identity}\end{equation}
 which comes from the unitarity and crossing symmetry of the bulk
S-matrix, the general reflection factor is equivalent to \[
P_{|0/1;n_{1},m_{1},\dots,n_{k-1},m_{k-1},\dots\rangle}^{+}=P_{|0/1\rangle}^{+}\prod_{k}\frac{a(\theta-i\nu_{n_{k}})a(\theta+i\nu_{n_{k}})}{a(\theta-i(\pi-w_{m_{k}}))a(\theta+i(\pi-w_{m_{k}}))}.\]

On physical grounds one expects that the DSG model on the strip with
two boundaries should have in general pairs of the DM type BBS in
the spectrum when $L\rightarrow\infty$. In this paper we consider
a somewhat simpler situation where the boundary parameters on one
boundary do not allow any DM bound state. In this case, one expects
that only one set of the DM BBSs are present in $L\rightarrow\infty$
limit. In the following we show how the solutions of the NLIE with
purely imaginary roots meet this expectation.

\section{Large volume behavior of the NLIE}

In this section we provide the interpretation of the boundary strings
of the NLIE in the large volume limit by mapping them to the boundary
bound states classified by DM. First we show that the asymptotic analysis
for the existence of a boundary string in NLIE is equivalent to the
BAE analysis, then we focus on their interpretation. For pedagogical
reasons we present the results for the repulsive regime first, where
we have at most one BBS and then turn to the more complicated problem
of the attractive regime. Finally we confirm our findings by calculating
the boundary Lüscher corrections for the ground-states.

\subsection{Large volume analysis of the boundary excited state NLIE}

The aim of this subsection is to replace the $\epsilon$ analysis
for the $(n,m)$ strings in the BAE by a source term analysis in the
infrared (IR) limit of the NLIE (\ref{NLIE}). In this analysis the
counting function $Z(\theta)$ is replaced by its asymptotic (large
volume) form: \begin{equation}
Z\left(\theta\right)=2ML\sinh\theta+P_{{\rm \mathrm{bdry}}}(\theta)-\sum\limits _{k}(\chi(\theta-\theta_{k})+\chi(\theta+\theta_{k})),\label{szamlalo1}\end{equation}
 and the quantization condition is obtained from \begin{equation}
e^{iZ(\theta_{j})}=1.\qquad j=1,\dots\label{szamlalo2}\end{equation}
 Using the relations (\ref{boundparams}) and (\ref{eq:xiphi}) the
boundary parameters $H_{\pm}$ are related to the DM $\xi_{\pm}$
parameters as \begin{equation}
H_{\pm}=p(1\mp\frac{2\xi_{\pm}}{\pi}).\label{xidef}\end{equation}
 Some care is required for Eqs.(\ref{szamlalo1},\ref{szamlalo2})
since one may have to use $\chi_{\mathrm{wide}}(\theta)$ instead
of $\chi(\theta)$ depending on the location of the roots and also
the second determination form of all the quantities. The expectation
is that from this asymptotic analysis one can obtain the same string
like objects as from the $\epsilon$ analysis in the BAE. This would
confirm the relevance of the results of the $\epsilon$ analysis,
as here we work with the (large volume limit of the) exact ground
state(s) as opposed to the pseudo-vacuum in the $\epsilon$ analysis.

In this analysis we keep $H_{-}$ in the domain where we expect no
bound state on this boundary ($0<H_{-}<2p$) while we let $H_{+}$
to move from a similar domain into $-1<H_{+}<0$ where bound states
are expected. Consider first the zero string case when in Eq.(\ref{szamlalo1})
there are just two source terms with $\pm\theta_{0}$ and assume $H_{+}$
is positive $0<H_{+}$ (i.e. $\xi_{+}<\pi/2$). Using the well known
identity \begin{equation}
e^{iP_{{\rm \mathrm{bdry}}}(\theta)}=-\frac{P_{\vert0\rangle}^{+}(\theta,\xi_{+})P_{\vert0\rangle}^{+}(\theta,\xi_{-})}{a(2\theta)}\label{PRRS}\end{equation}
 where $P_{\vert0\rangle}^{+}(\theta,\xi)$ is the Ghoshal-Zamolodchikov
ground state soliton reflection amplitude (\ref{Intreps}) and $a(\theta)$
denotes the bulk soliton-soliton scattering the only equation in Eq.(\ref{szamlalo2})
becomes: \[
e^{i2ML\sinh\theta_{0}}\frac{P_{\vert0\rangle}^{+}(\theta_{0},\xi_{+})P_{\vert0\rangle}^{+}(\theta_{0},\xi_{-})}{a(2\theta_{0})a(2\theta_{0})}=1.\]
 An imaginary root corresponding to a BBS would show up in the form
of a solution $\theta_{0}=iv_{0}+\epsilon$ with $v_{0}$ in the physical
domain ($0<v_{0}<\frac{\pi}{2}$) and $\epsilon\rightarrow0$ for
$L\rightarrow\infty$. This can happen only if $iv_{0}$ is a pole
of one of the $P^{+}$'s. However, they have no poles in the physical
strip when both $\xi_{\pm}<\frac{\pi}{2}$. (Note that the $a$'s
in the denominator cancel also the boundary independent poles of the
two $P^{+}$'s). Thus for both $H_{\pm}$ positive this asymptotic
analysis gives no hint of a bound state. This is also consistent with
the counting equation (\ref{conticount}) that for $H_{\pm}>0$ and
$N_{H}=0$ allows only a solution with $M_{c}=0$.

Now let $H_{+}$ become negative, but still consider the zero string
case since the counting equation now allows $M_{c}=1$. The crucial
observation is that the exponential of $iP_{{\rm \mathrm{bdry}}}(\theta)$
contains in this case the excited state soliton reflection amplitude
$P_{|1;0\rangle}^{+}(\theta,\xi_{+})$: \[
e^{iP_{{\rm \mathrm{bdry}}}(\theta)}|_{-1<H_{+}<0}=-\frac{P_{|1;0\rangle}^{+}(\theta,\xi_{+})P_{\vert0\rangle}^{+}(\theta,\xi_{-})}{a(2\theta)}.\]
 In DM \cite{MD} $P_{|1;0\rangle}^{+}(\theta,\xi_{+})$ is expressed
in two equivalent ways: \[
P_{|1;0\rangle}^{+}(\theta,\xi_{+})=P_{\vert0\rangle}^{+}(\theta,\xi_{+})a(\theta-\nu_{0})a(\theta+\nu_{0}),\quad P_{|1;0\rangle}^{+}(\theta,\xi_{+})=\overline{P_{\vert0\rangle}^{-}(\theta,\xi_{+})}=P_{\vert0\rangle}^{+}(\theta,\xi_{+}-\frac{\pi}{p}-\pi).\]
 The first form is natural from the bootstrap point of view and makes
it easy to see that $P_{|1;0\rangle}^{+}(\theta,\xi_{+})$ has poles
at $i\nu_{0}$ and at $i\nu_{-N}$ for $N=1,2,\dots$, while the second
form (where the over-line describes the transformation $\xi_{+}\rightarrow\pi(1+p^{-1})-\xi_{+}$)
is useful to verify the integral representation. To support the claim
we write here $F(\theta,H_{+})$ for $H_{+}>0$: \[
F(\theta,H_{+})=\int_{-\infty}^{\infty}\frac{dk}{2\pi}\mathrm{e}^{ik\theta}\frac{\sinh({kp}(\xi_{+}+\frac{\pi}{2p})) }{2\sinh\frac{p\pi}{2}k\cosh\frac{\pi}{2}k},\]
 and for $H_{+}<0$: \[
F(\theta,H_{+})=\int_{-\infty}^{\infty}\frac{dk}{2\pi}\mathrm{e}^{ik\theta}\frac{\sinh({kp}(\xi_{+}-\pi-\frac{\pi}{2p}))}{2\sinh\frac{p\pi}{2}k\cosh\frac{\pi}{2}k},\]
 showing that they are indeed connected by the transformation in the
second DM form. Note that this implies that in this domain of $H_{\pm}$
the integral equation describes the excited state $|1;0\rangle$.

Using this observation in Eq.(\ref{szamlalo2}) leads to the quantization
condition \begin{equation}
e^{i2ML\sinh\theta_{0}}\frac{P_{|1;0\rangle}^{+}(\theta_{0},\xi_{+})P_{\vert0\rangle}^{+}(\theta_{0},\xi_{-})}{a(2\theta_{0})a(2\theta_{0})}=1.\label{egypol1}\end{equation}
 Since $P_{|1;0\rangle}^{+}(\theta,\xi_{+})$ has poles in the physical
strip this equation admits a bound state solution \[
\theta_{0}=i(\nu_{0}+\epsilon)\qquad{\textrm{with}}\qquad\epsilon\sim Re^{-2ML\sin\nu_{0}},\]
 (where $iR$ is the residue of the pole at $i\nu_{0}$) satisfying
the requirements described earlier. Eq.(\ref{egypol1}) is correct
if $\theta_{0}$ is in the first determination; but this condition
is met in a domain where $\xi_{+}$ just exceeds $\pi/2$ ($H_{+}$
is just below $0$) both in the $p>1$ (repulsive) and in the $p<1$
(attractive) domains. Even for these $\xi_{+}$-s the poles of $P_{|1;0\rangle}^{+}(\theta,\xi_{+})$
at $i\nu_{-N}$ are in the second determination thus cannot be used
to find solutions to Eq.(\ref{egypol1}) since the form of the equation
changes there. Furthermore in the repulsive regime also the counting
equation would require to introduce something else (possibly moving
objects) to compensate the presence of the wide roots.

If $\xi_{+}$ exceeds $3\pi/2$ then also $i\nu_{0}$ gets into second
determination ($\nu_{0}>p\pi$) and we have to reconsider the asymptotic
analysis and the solution we found even for the zero string case.
(Since $\xi_{+}\leq\xi_{{\rm max}}=\frac{\pi}{2}(1+p^{-1})$, $\xi_{+}=3\pi/2$
is in this allowed range only if $p<1/2$ in the attractive domain).
In the quantization condition, Eq.(\ref{szamlalo2}), now $Z(\theta_{0})_{II}$
appears, where \[
Z\left(\theta_{0}\right)_{II}=2ML(\sinh\theta_{0}-\sinh(\theta_{0}-ip\pi))+P_{{\rm \mathrm{bdry}}}(\theta_{0})-P_{{\rm \mathrm{bdry}}}(\theta_{0}-ip\pi)-{\textrm{source}}\]
 with \[
{\textrm{source}}=(\chi(2\theta_{0})+\chi(2\theta_{0}-2ip\pi)-2\chi(2\theta_{0}-ip\pi)).\]
 As a consequence Eq.(\ref{szamlalo2}) now takes the form: \begin{equation}
e^{i2ML(\sinh\theta_{0}-\sinh(\theta_{0}-ip\pi))}\frac{P_{|1;0\rangle}^{+}(\theta_{0},\xi_{+})P_{\vert0\rangle}^{+}(\theta_{0},\xi_{-})}{a(2\theta_{0})a(2\theta_{0})}\frac{(a(2\theta_{0}-ip\pi))^{2}}{P_{|1;0\rangle}^{+}(\theta_{0}-ip\pi,\xi_{+})P_{\vert0\rangle}^{+}(\theta_{0}-ip\pi,\xi_{-})}=1.\label{egypol2}\end{equation}
 Note that the $P_{|1;0\rangle}^{+}$ in the denominator cancels all
but the $i\nu_{0}$ pole of $P_{|1;0\rangle}^{+}$ in the numerator,
thus Eq.(\ref{egypol2}) admits only a bound state solution of the
form \[
\theta_{0}=i(\nu_{0}+\epsilon)\qquad{\textrm{with}}\qquad\epsilon\sim Re^{-2ML(\sin\nu_{0}-\sin(\nu_{0}-p\pi))}.\]
 Thus the asymptotic analysis gives a possibility for a zero string
bound state solution if $H_{+}<0$ independently whether $i\nu_{0}$
is in the first or in the second determination. However, since the
source terms are different in the two cases ($\chi\rightarrow\chi_{\mathrm{wide}}$
for $i\nu_{0}$ in the second determination), the interpretation of
the bound states is different: while in the first case it corresponds
to the ground state, in the second it corresponds to the state $|1;1\rangle$
as described in subsection 3.2.

Let's now turn to longer strings that may appear only in the attractive
regime, where wide roots can be added freely to the NLIE since their
number cancels from the counting equation (\ref{conticount}). To
describe how these string like structures appear form Eq.(\ref{szamlalo2})
note that most of the imaginary roots of the string are in the second
determination thus Eq.(\ref{szamlalo2}) takes the form \[
Z(\theta)=2ML\sinh\theta+P_{{\rm \mathrm{bdry}}}(\theta)-\sum_{j=n}^{-m}\left(\chi_{II}^{-}(\theta-\theta_{j})+\chi_{II}^{+}(\theta+\theta_{j})\right)\]
 The location of the roots are $\theta_{j}=i\nu_{j}+i\epsilon_{j}(L)$,
where we suppose that $\epsilon_{j}(L)\to0$ as $L\to\infty$. If
$\Im m(\theta_{n})<p\pi$ then the term corresponding to $j=n$ in
the sum is replaced with $\chi(\theta-\theta_{n})+\chi(\theta+\theta_{n})$.
The other terms are defined as $f_{II}^{\pm}(\theta)=f(\theta)-f(\theta\mp i\pi p)$.
We also suppose that $\Im m(\theta_{-m})<\pi$ since we need a BBS
state with non-vanishing energy.

The position of the roots is determined by the quantization condition
$e^{iZ_{II}^{+}(\theta_{k})}=1$ for wide roots and $e^{iZ(\theta_{n})}=1$
for the close root if there is one. The condition can be written as
\begin{equation}
e^{2iML\sinh_{II}^{+}(\theta_{k})}\frac{e^{iP_{{\rm \mathrm{bdry}\, II}}^{+}(\theta_{k})}}{a(2\theta_{k})}\frac{a(2\theta_{k}-i\pi p)^{2}}{a(2\theta_{k}-2i\pi p)}\prod_{j\neq k}\frac{1}{a_{II}^{+}(\theta_{k}+\theta_{j})_{II}^{+}\: a_{II}^{-}(\theta_{k}-\theta_{j})_{II}^{+}}=1\label{szamlalo3}\end{equation}
Since $e^{-2ML\sin\nu_{k}}\to0$ in the limit $L\to\infty$ we have
to analyze the singularity structure of the function appearing in
Eq.(\ref{szamlalo3}). The analysis of the zero string case showed
that $\frac{e^{iP_{{\rm \mathrm{bdry}\, II}}^{+}(\theta_{k})}}{a(2\theta_{k})}$
has a pole at $\theta=i\nu_{0}$. Furthermore \[
a_{II}^{-}(\theta)_{II}^{+}=\frac{a^{2}(\theta)}{a(\theta+i\pi p)a(\theta-i\pi p)}\]
 has a pole at $\theta=i\pi p$ and a zero at $\theta=-i\pi p$. Focusing
on the divergent terms in Eq.(\ref{szamlalo3}) one obtains the equations:
\begin{eqnarray*}
 & e^{(\lambda_{n+1}-\lambda_{n})L}\propto\epsilon_{n}-\epsilon_{n-1}\;;\quad e^{(\lambda_{n}-\lambda_{n-1})L}\propto\frac{\epsilon_{n-1}-\epsilon_{n-2}}{\epsilon_{n}-\epsilon_{n-1}}\;;\quad\dots\;;\quad e^{(\lambda_{k}-\lambda_{k-1})L}\propto\frac{\epsilon_{k-1}-\epsilon_{k-2}}{\epsilon_{k}-\epsilon_{k-1}}\;;\quad\dots\\
 & e^{(\lambda_{2}-\lambda_{1})L}\propto\frac{\epsilon_{1}-\epsilon_{0}}{\epsilon_{2}-\epsilon_{1}}\;;\quad e^{(\lambda_{1}-\lambda_{0})L}\propto\epsilon_{0}\frac{\epsilon_{0}-\epsilon_{-1}}{\epsilon_{1}-\epsilon_{0}}\;;\quad e^{(\lambda_{0}-\lambda_{-1})L}\propto\frac{\epsilon_{-1}-\epsilon_{-2}}{\epsilon_{0}-\epsilon_{-1}}\;\dots\,;\,\, e^{(\lambda_{-m+1}-\lambda_{-m})L}\propto\frac{1}{(\epsilon_{-m+1}-\epsilon_{-m})}\end{eqnarray*}
 where $\lambda_{k}=2M\sin\nu_{k}$ and whenever $\theta_{n}$ is
a close root then in the first equation we have to put $\lambda_{n+1}=0$.

Since all the $\epsilon$-s have to go to zero as $L\rightarrow\infty$
these equations lead to the following requirements: \begin{eqnarray*}
\lambda_{n+1}<\lambda_{n}\quad;\qquad & \lambda_{n+1}<\lambda_{n-1} & \;;\quad\dots\;;\quad\lambda_{n+1}<\lambda_{1}\\
 & \lambda_{-m}>\lambda_{n+1}\\
\lambda_{-m}<\lambda_{-m+1}\quad;\qquad & \lambda_{-m}<\lambda_{-m+2} & \;;\quad\dots\;;\quad\lambda_{-m}<\lambda_{0}\end{eqnarray*}
 The strongest condition from the first line is $\lambda_{n+1}<\lambda_{1}$
which means $\nu_{1}-\frac{\pi}{2}<\frac{\pi}{2}-\nu_{n+1}$. The
second requirement can be translated to $\nu_{-m}-\frac{\pi}{2}<\frac{\pi}{2}-\nu_{n+1}$
which is equivalent to $\nu_{n+1}<\pi-\nu_{-m}=w_{m}$. Finally the
strongest condition from the last line is $\lambda_{-m}<\lambda_{0}$
which is equivalent to $w_{m}<\nu_{0}$. These last two conditions
are completely equivalent to the ones Saleur-Skorik obtained for the
$(n,m)$ strings from the BAE on the one hand, while they are also
consistent with the DM bounds (\ref{MattsonDorey}) on the other.

\subsection{Bound-state NLIE in the repulsive regime}

In this subsection the interpretation of the BBS in the repulsive
regime is elaborated. The validity range of the pure NLIE -- Eq.(\ref{NLIE})
without source terms -- as it is derived from the BAE using Fourier
transformation, is $-2p-2<H_{\pm}<2p+2$. We give its interpretation
in this full range. The symmetry $H_{\pm}\to H_{\pm}+2p+2$ of the
BAE survives at the NLIE level (can be checked also by explicit comparison),
thus the $[0,2p+2]$ domain is equivalent to the $[-2p-2,0]$ domain.
One possibility to compare with the DM spectrum is to put a particle
between the two boundaries and analyze its reflection factors by comparing
the large volume limit of the NLIE with the Bethe-Yang quantization
condition. We recall this analysis from \cite{DNLIE} here.

In the repulsive regime in the half line theory we have at most one
BBS and its energy can be plotted as function of the boundary parameter
$\xi$ as shown on Figure 2.

\begin{figure}[bh]
\begin{centering}
\includegraphics[width=10cm]{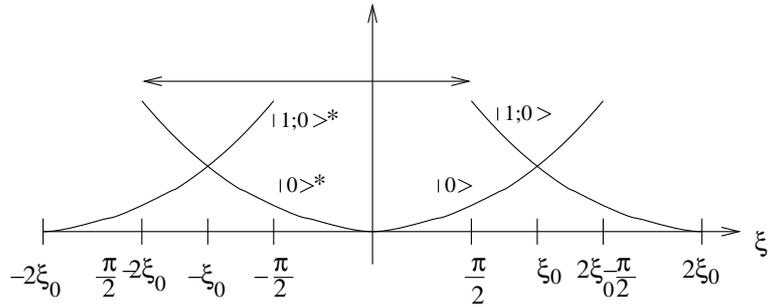} 
\par\end{centering}

\caption{Shematic figure of the boundary condition dependent part of the energies
of the various ground-states }

\end{figure}

Figure 2 is schematic and shows only the $\xi$ dependent part of
the boundary energies of the ground-state, $E_{\vert0\rangle}(\xi)=-\frac{M\cos p\xi}{2\cos\frac{\pi p}{2}}$
\cite{LMSS}, and the BBS, $E_{\vert1;0\rangle}=E_{\vert0\rangle}+\cos\nu_{0}$.
The discrete symmetries $\xi\to-\xi$ and $\xi\to2\xi_{0}-\xi$ induce
maps between the states $\vert0\rangle\to\vert0\rangle^{*}$ and $\vert0\rangle\to\vert1;0\rangle$
but in the same time the soliton has to be exchanged with the anti-soliton.
The two theories $\xi$ and $-\xi$ are not equivalent so we distinguish
their states by star. The transformation $\xi\to\xi+2\xi_{0}$ maps
$\vert1;0\rangle^{*}\to\vert0\rangle$ and $\vert0\rangle^{*}\to\vert1;0\rangle$
without changing the species. These transformations correspond to
the $\phi\to-\phi,$ $\phi\to\frac{2\pi}{\beta}-\phi$ and $\phi\to\phi+\frac{2\pi}{\beta}$
transformations of the Lagrangian and, by means of them, the parameter
range of the theory can be restricted to $\phi\in[0,\frac{\pi}{\beta}]$
or equivalently to $\xi\in[0,\xi_{0}]$.

The boundary condition dependent part of the reflection factor of
the soliton on the ground-state boundary can be written as (\ref{Intreps})
which is valid in the domain $0<\xi<\frac{\pi}{2}$. The bound at
$\frac{\pi}{2}$ signals the pole of the reflection factor which corresponds
to the boundary state $\vert1;0\rangle$. The integral representation
is valid also for $-\frac{\pi}{2}(1+2p^{-1})<\xi<0$ but here it corresponds
to the state $\vert0\rangle^{*}$. The validity range of the integral
representation is marked with the arrow on the figure.

Now putting one hole into the pure NLIE for $2p>H_{\pm}>0$, and comparing
the large volume limit of the quantization condition to the Bethe-Yang
equation \[
e^{iZ(\theta_{1})}=1\qquad\longleftrightarrow\qquad e^{2iML\sinh\theta_{1}}e^{iP_{\vert0\rangle}(\xi_{+},\theta_{1})}e^{iP_{\vert0\rangle}(\xi_{-},\theta_{1})}=1\]
 one arrives at the identifications of the parameters (\ref{xidef}).
Let us fix $H_{-}$ in the domain $[2p,0]$ such that this boundary
does not allow any bound-state and scan the whole $-2p-2<H_{+}<2p+2$
range on the other boundary. The previous findings show that the integral
equation in the domain $H_{+}\in[0,p]$ describes the ground state
$\vert0\rangle$. They also show that in the range $H_{+}\in[p,2p+2]$
it describes the $\vert0\rangle^{*}$ state instead. This is interesting
since for $H_{+}\in[2p,2p+1]$ there exists a BBS in the spectrum
but the pure NLIE corresponds to the ground-state. The same NLIE in
the $H_{+}\in[2p+1,2p+2]$ domain describes the BBS, and using the
symmetry $H\to H+2p+2$ we can conclude that it corresponds also to
the excited boundary state $\vert1;0\rangle$ in the $H_{+}\in[-1,0]$
range. For $H_{+}\in[-2-p,-1]$ it describes the $\vert1;0\rangle$
ground-state. This can also be confirmed by comparing the reflection
factor of the soliton on the state $\vert1;0\rangle$ to the result
coming from the NLIE for $H_{+}<0$. 

In summarizing using the reflection factors of the solitonic states
we conclude that the pure NLIE describes the state marked with $+$
for $H_{+}>0$ and the one marked with $-$ for $H_{+}<0$ in Figure
3.

\begin{figure}[h]
\begin{centering}
\includegraphics[width=10cm]{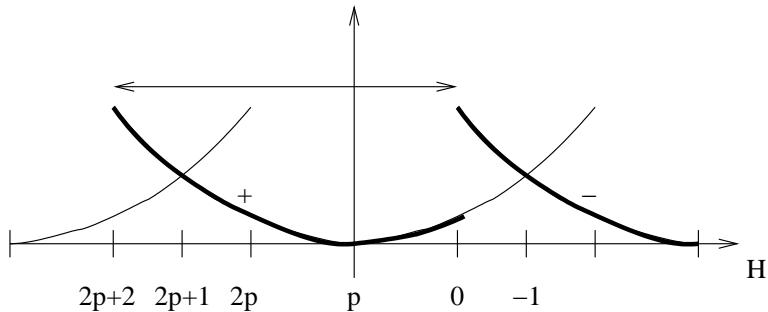}
\par\end{centering}

\caption{The pure NLIE describes the state marked with + }

\end{figure}

The boundary energies cannot be calculated either from the NLIE or
from the bootstrap since in both approaches the ground-state energy
is normalized to $0$ and energy differences can be determined only.

Let us introduce a boundary imaginary root in the NLIE (\ref{NLIE}).
The BAE and the asymptotic analysis of the NLIE predict the position
of the root to be \[
\theta=iu\approx i\frac{\pi}{2}(n(2p+2)-H_{+})\quad,\qquad n\in Z\]
 which we plot on Figure 4.  

\begin{figure}
\begin{centering}
\includegraphics[width=10cm]{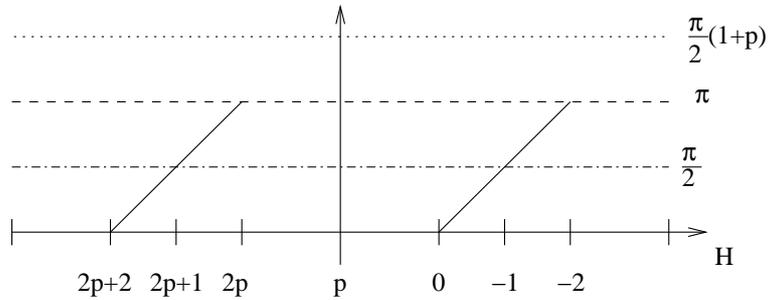}
\par\end{centering}

\caption{The straight lines show the location of the imaginary root as a function
of $H$}

\end{figure}

\noindent On Figure 4 the first determination is below $i\pi$ marked
with a dashed line while the self-conjugate line indicated by a dotted
line is at $i\frac{\pi}{2}(1+p)$. Exactly when $H_{+}=2p$ the $n=1$
imaginary root appears at $i\pi$. Adding it to the NLIE via the source
term $-\chi(\theta-iu)-\chi(\theta+iu)$ modifies the reflection factor
and increases the energy by $-M\cos u$. Clearly this is positive
for $H_{+}<2p+1$ zero for $H_{+}=2p+1$, negative between $2p+1<H_{+}<2p+2$
and leaves the imaginary axis at $H=2p+2$. So from comparing the
energy differences between the pure NLIE and the NLIE with the imaginary
root added we suspect that the later one describes the state denoted
by the dotted line on Figure 5. 

\noindent %
\begin{figure}[h]
\begin{centering}
\includegraphics[width=10cm]{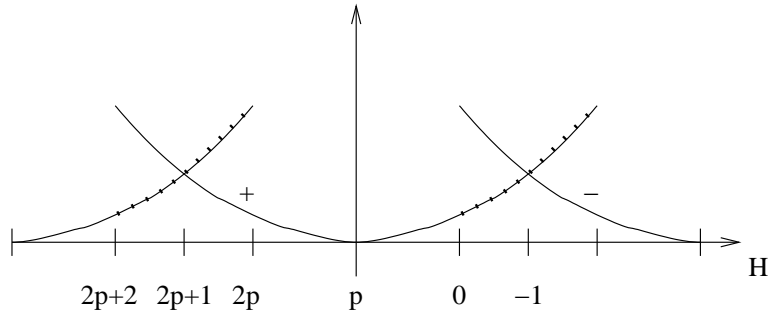}
\par\end{centering}

\caption{The NLIE with the imaginary root added describes the state marked
with the dotted line}

\end{figure}

\noindent This can be confirmed also by putting one additional hole
and analyzing the reflection factors, which we show below.

Since all these states can be described in the $p>H_{+}>-1$ or equivalently
in the $0<\xi_{+}p<\frac{\pi}{2}(1+p)$ domain we focus on this from
now on and summarize the previous findings. The pure NLIE describes
the ground-state $|0\rangle$ for the range $p>H_{+}>0$, while for
the range $0>H_{+}>-1$ it describes the state $|1;0\rangle$. If
we, in this range, add an imaginary root located at $iu_{0}$ (where
$u_{0}=\nu_{0}=p\xi-\frac{\pi p}{2}$) we change the energy by $-M\cos\nu_{0}$
and the logarithm of the reflection factor by $-\chi(u-i\nu_{0})-\chi(u+i\nu_{0})$.
The energy of this state is then \[
E=E_{|1;0\rangle}-M\cos\nu_{0}\]
 while the reflection factor is \[
P^{+}(\theta)=P_{|1;0\rangle}^{+}(\theta)\frac{1}{a(\theta-i\nu_{0})a(\theta+i\nu_{0})}\]
 from these two expressions we can read off (\ref{exref}) that the
state with the imaginary roots at $\pm i\nu_{0}$ added is the ground
state $|0\rangle$.

We can argue another independent way that the state with the root
added in the $H_{+}\in[0,-1]$ range describes the ground-state. We
can perform an analytic continuation of the pure NLIE from the $H_{+}>0$
domain. This method is standard and was used frequently to obtain
the excited states in the TBA equations \cite{YL,Chaiho,BRZ}. By
changing the sign of $H_{+}$ two singularities of $\log(1-e^{iZ})$
are crossing the contours and by encircling them and picking up the
residue terms the NLIE with one root pair added can be obtained from
the pure NLIE.

\subsection{Bound-state NLIE in the attractive regime}

In this subsection we map the large volume limit of the boundary strings
to the BBSs of DM in the attractive regime. Since all states can be
described in the $p>H_{+}>-1$ domain we concentrate only on this
range. First we figure out the correspondence from the boundary energies
and then confirm our findings by comparing the solitonic reflection
factors, too.

From the bootstrap point of view the properties of the first excited
boundary state $\vert1;0\rangle$, such as energy and reflection factor,
are the same as in the repulsive regime and the interpretation of
the pure NLIE is completely analogous: for $H_{+}>0$ it describes
the ground-state $\vert0\rangle$, while for $-1<H_{+}<0$ it corresponds
to the BBS $\vert1;0\rangle.$ In the attractive regime, however,
where breathers are also in the spectrum we can confirm this assignment
independently by analyzing the large volume behavior of the first
breather.

In doing so we insert a self-conjugate root into the large volume
limit of the pure NLIE (\ref{NLIE})\[
Z\left(\theta\right)=2ML\sinh\theta+P_{{\rm \mathrm{bdry}}}(\theta)-(\chi_{II}(\theta-\alpha)+\chi_{II}(\theta+\alpha))\quad;\qquad\alpha=\theta_{0}+i\frac{\pi}{2}(p+1)\]
 and compare the $e^{iZ(\alpha)_{II}}=1$ quantization condition to
the first breather's Bethe-Yang equation \begin{equation}
e^{i2m_{1}L\sinh\theta_{0}}R_{H_{+}}^{(1)}(\theta_{0})R_{H_{-}}^{(1)}(\theta_{0})=1\label{br1BY}\end{equation}
 Here $m_{1}=2M\sin\frac{\pi p}{2}$ is the mass of the first breather
and\[
R_{H}^{(1)}(\theta)=\frac{\left(2+p\right)_{\theta}\left(1\right)_{\theta}}{\left(3+p\right)_{\theta}}\left[\frac{\left(p-|H|-1\right)_{\theta}}{\left(p-|H|+1\right)_{\theta}}\right]^{\mathrm{sign}(H)},\quad(x)_{\theta}=\frac{\sinh(\frac{\theta}{2}+i\frac{\pi x}{4})}{\sinh(\frac{\theta}{2}-i\frac{\pi x}{4})}\]
 denotes its reflection factor \cite{Ghoshal}. Using the integral
representation of the combination \[
(x)_{\theta+i\frac{\pi}{2}}(x)_{-\theta+i\frac{\pi}{2}}=\exp\left[\int_{-\infty}^{\infty}\frac{dt}{t}e^{\frac{it\theta}{\pi}}\frac{\sinh t(1-\frac{x}{2})}{\cosh(t/2)}\right]\]
 together with the identity $\left(1+p\right)_{i\frac{\pi}{2}\mp\theta}=(1-p)_{i\frac{\pi}{2}\pm\theta}$
one can indeed map the quantization condition (\ref{szamlalo2}) to
the Bethe-Yang equation (\ref{br1BY}). Furthermore, from the first
breather's reflection amplitude emerging from this comparison one
can see that the pure NLIE describes the ground-state $\vert0\rangle$
for $H_{+}>0$, while for $H_{+}<0$ it gives the BBS $\vert1;0\rangle$.

Once we know the interpretation of the pure NLIE we turn to the analysis
of the $(n,m)$ string allowed by both the BAE and the asymptotic
analysis of the NLIE. We include the following source term in the
NLIE \[
\mbox{source}=-\sum_{j=-n+1}^{m}(\chi_{II}(\theta-iu_{j})+\chi_{II}(\theta+iu_{j}))-\left\{ \begin{array}{c}
\chi_{II}(\theta-iu_{-n})+\chi_{II}(\theta+iu_{-n})\quad\textrm{if}\quad u_{-n}>\pi p\\
\chi(\theta-iu_{-n})+\chi(\theta+iu_{-n})\quad\textrm{if}\quad u_{-n}<\pi p\end{array}\right.\]
 and at the same time make the corresponding modification of the energy
\[
E=-M\sum_{j=-n+1}^{m}\cos_{II}(u_{j})-\left\{ \begin{array}{c}
\cos_{II}(u_{-n})\quad\textrm{if}\quad u_{-n}>\pi p\\
\cos(u_{-n})\quad\textrm{if}\quad u_{-n}<\pi p\end{array}\right.\]
 The contribution of the wide and close roots reads explicitly \[
\mbox{source}=-\chi(\theta-iu_{m})-\chi(\theta+iu_{m})\quad;\qquad E=-M\cos u_{m}\]
 if $u_{-n}<p\pi$ and \[
\mbox{source}=-\chi(\theta-iu_{m})-\chi(\theta+iu_{m})+\chi(\theta-iu_{-n-1})+\chi(\theta+iu_{-n-1})\quad;\qquad E=-M\cos u_{m}+M\cos u_{-n-1}\]
 if $u_{-n}>p\pi$. The NLIE and the BAE is periodic with period $i\pi(1+p)$.
From now on we use the imaginary strip between $0$ and $\pi(1+p)$
as the fundamental range in contrast to the usual $[-\frac{\pi}{2}(p+1),\frac{\pi}{2}(p+1)]$.
Roots with negative imaginary parts are mapped to the upper half plane
by the $u\to u+\pi(1+p)$ transformation as is demonstrated on Figure
6.

\begin{figure}[h]
\begin{centering}
\includegraphics[height=6cm,keepaspectratio]{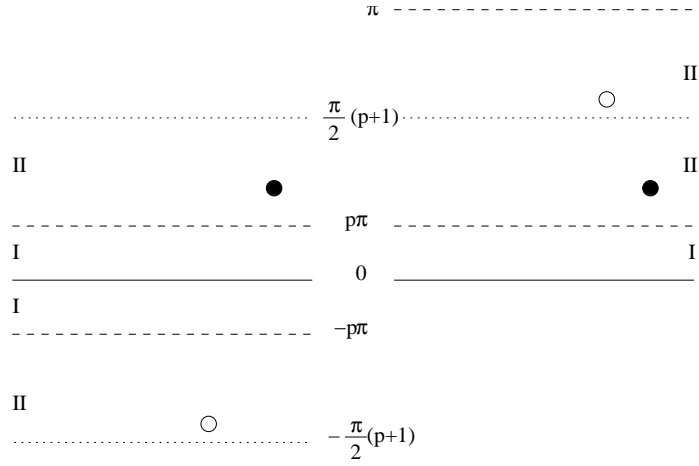}
\par\end{centering}

\caption{The two fundamental strips used to describe imaginary roots }

\end{figure}

Taking any allowed root $-\pi p>\bar{u}>-\frac{\pi}{2}(p+1)$ we replace
it with the root $u=\bar{u}+\pi(p+1)$ which is now in the strip $\frac{\pi}{2}(p+1)<u<\pi$.
In the energy formula the corresponding change is to replace $\cos_{II}(\bar{u})$
by $\cos_{II}(u)$, but they are equal since \[
\cos_{II}(\bar{u})=\cos(\bar{u})-\cos(\bar{u}+\pi p)=\cos(u-\pi p-\pi)-\cos(u-\pi)=\cos u-\cos(u-\pi p)=\cos_{II}(u)\]
 Similarly in the source term we replace $\chi_{II}(\theta-i\bar{u})+\chi_{II}(\theta+i\bar{u})$
by $\chi_{II}(\theta-iu)+\chi_{II}(\theta+iu)$. Their equality follows
from the fact that $\chi_{II}(\theta-iu)+\chi_{II}(\theta+iu)$ is
symmetric for $u=\frac{\pi}{2}(p+1)$,which is a consequence of the
identity (\ref{eq:identity}).

Let's analyze now the boundary energy as well as the reflection factors
of the solitons in case of the $(n,m)$ string added. To make correspondence
with the BAE we remark that $\nu_{0}=u_{0}=\xi p-\frac{\pi p}{2}$.
As a consequence \[
u_{-n}=\nu_{n}\quad;\qquad u_{m}=\pi-w_{m}\]

We start the analysis with the simplest $m=0$ string. We have to
distinguish two cases as shown on Figure 7. 

\begin{figure}[h]
\begin{centering}
\includegraphics[height=4cm]{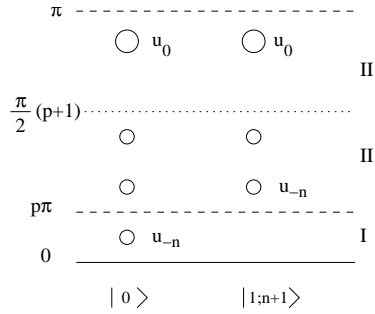}
\par\end{centering}

\caption{The figure of the two possible $(n,0)$ strings}

\end{figure}

In the first case (left) $u_{-n}<\pi p$. The energy compared to the
state $|1;0\rangle$ (described by the pure NLIE) is \[
E=E_{|1;0\rangle}-M\sum_{j=n}^{0}\cos_{II}\nu_{j}-M\cos\nu_{n}=E_{|1;0\rangle}-M\cos\nu_{0}\]
 while for the reflection factor we obtain the factors \[
P^{+}(\theta)=P_{|1;0\rangle}^{+}a(\theta-i\nu_{0})^{-1}a(\theta+i\nu_{0})^{-1}\]
 This is the same result we obtained in the repulsive regime when
added one root at $i\nu_{0}$ thus we conclude that it describes the
ground-state $|0\rangle$. So the ground-state corresponds to the
longest string and this is true as far as $0<\nu_{0}<\frac{\pi}{2}$
which is equivalent to $0>H_{+}>-1$. This result was obtained also
by \cite{Salsko}. Let us note that we can describe the ground-state
by analytic continuation in $H_{+}$ as we did in the repulsive case.
The difference being, that once $H_{+}$ reaches the value when $\nu_{n}(H_{+})$
enters the physical strip we have to move it through the integration
contour which results in its source term. That is why the longest
possible $(0,n)$ string gives the vacuum.

Suppose now that $u_{-n}>\pi p$ (right on the figure) so the bottom
root is in the second determination, too. Using the second determination
of the cosine function we obtain the energy as \[
E=E_{|1;0\rangle}-\cos\nu_{0}+\cos\nu_{n+1}\]
 while the reflection factor turns out to be \[
P^{+}(\theta)=P_{|1;0\rangle}^{+}\frac{a(\theta-i\nu_{n+1})a(\theta+i\nu_{n+1})}{a(\theta-i\nu_{0})a(\theta+i\nu_{0})}\]
 This is all consistent with the proposal that this string corresponds
to the state $|1;n+1\rangle.$ From the DM analysis we know that this
state exists when $\nu_{n+1}>0$ which is just the statement $u_{-n}>\pi p$
we obtained from the asymptotic analysis.

Consider now the most general $(n,m)$ string with $m>0$. Distinguish
again two cases depending on whether $u_{-n}<\pi p$ or $u_{-n}>\pi p$
as shown on Figure 8.

\begin{figure}[h]
\begin{centering}
\includegraphics[height=6cm,keepaspectratio]{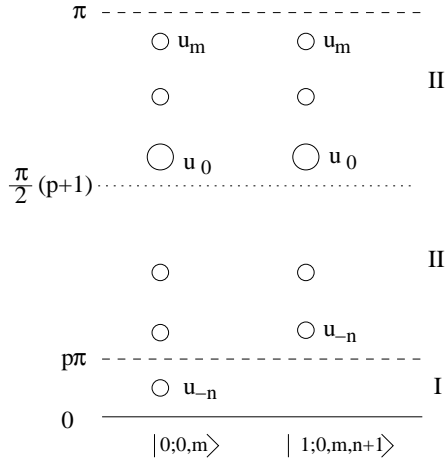}
\par\end{centering}

\caption{The figure of the two possible $(n,m)$ strings}

\end{figure}

In the first case (left) $u_{-n}<\pi p$ the energy of an $(n,m)$
string is \[
E=E_{|1;0\rangle}-\cos u_{m}=E_{|1;0\rangle}+\cos w_{m}\]
 while the reflection factor is \begin{eqnarray*}
P^{+}(\theta) & = & P_{|1;0\rangle}^{+}\frac{1}{a(\theta+i(\pi-w_{m}))a(\theta-i(\pi-w_{m}))}\\
 & = & P_{|0\rangle}^{+}\frac{a(\theta-i\nu_{0})a(\theta+i\nu_{0})}{a(\theta+i(\pi-w_{m}))a(\theta-i(\pi-w_{m}))}\end{eqnarray*}
 Both the energy and the reflection factor is consistent with the
identification of the string with the $|0;0,m\rangle$ BBS. From the
DM bootstrap analysis we know that this bound-state exists whenever
$\frac{\pi}{2}>\nu_{0}>w_{m}>0$. The $\nu_{0}>w_{m}$ condition is
equivalent to condition $u_{1}+u_{m}>\pi(p+1)$ as obtained from the
asymptotic analysis. The $w_{m}>0$ condition is equivalent to $u_{m}<\pi$
which follows from the continuum counting equation (\ref{conticount}).

For $u_{-n}>\pi p$ the energy of the $(n,m)$ string is \[
E=E_{|1;0\rangle}+\cos w_{m}+\cos\nu_{n+1}\]
 while the reflection factor is \[
P^{+}(\theta)=P_{|1;0\rangle}^{+}\frac{a(\theta+i\nu_{n+1})a(\theta-i\nu_{n+1})}{a(\theta+i(\pi-w_{m}))a(\theta-i(\pi-w_{m}))}\]
 From which we can conclude that the corresponding state is $|1;0,m,n+1\rangle.$
This state exists whenever $\nu_{0}>w_{m}>\nu_{n+1}>0$. The new condition
compared to the previous discussions is $w_{m}>\nu_{n+1}$ but this
is equivalent to $u_{-n}+u_{m}<\pi(p+1)$. So this state exists exactly
the same time when the corresponding $(n,m)$ string in the BAE.

Suppose now that to the $(n_{1},m_{1})$ string we have already described
we add another $(n_{2},0)$ string with $n_{2}<n_{1}$. Since $u_{-n_{2}}>\pi p$
the second string increases the energy by $-M\cos\nu_{0}+M\cos\nu_{n_{2}+1}$
so changes the zero label to $n_{2}+1$. Explicitly if the original
state was $|0;0,m_{1}\rangle$ then the new state is $|0;n_{2}+1,m_{1}\rangle$
if, however, the original was $|1;0,m_{1},n_{1}+1\rangle$ then the
new state is $|1;n_{2}+1,m_{1},n_{1}\rangle$. If additionally to
the $(n_{1},m_{1})$ string we add another $(n_{2},m_{2})$ string
with $n_{2}<n_{1}$ and $m_{2}<m_{1}$,then, since $u_{-n_{2}}>\pi p$,
the second string increases the energy by $\cos w_{m_{2}}+\cos\nu_{n_{2}+1}$
so adds two labels. Concretely if the original state was $|0;0,m_{1}\rangle$
then the new state is $|0;0,m_{2},n_{2}+1,m_{1}\rangle$, if , however,
it was $|1;0,m_{1},n_{1}+1\rangle$ then the new state is $|1;0,m_{2},n_{2}+1,m_{1},n_{1}+1\rangle$.We
have checked explicitly that the energy formulas and the reflection
factors are consistent with these assumptions. For the existence of
this state the bootstrap gives the relation \emph{$\nu_{n_{2}+1}>w_{m_{1}}$}but
we were not able to find its analogue on the BAE side.

\subsection{Finite size correction of the ground-state energy: boundary Lüscher
correction}

In this subsection the large volume asymptotic of the ground-state
NLIE{\small{} }is analyzed and compared to the Lüscher type correction
\cite{BLusch}. The general form of this correction, valid in any
two dimensional boundary quantum field theory, was determined in \cite{BLusch}
and first we concretize the results for the sine-Gordon model with
Dirichlet boundary condition.

In the repulsive regime ($p>1$), where there is no breather in the
spectrum, the finite size energy correction in leading order is governed
by the soliton/anti-soliton reflection contribution as \[
E_{0}(L)=E_{0}(\infty)-M\int_{-\infty}^{\infty}\frac{d\theta}{4\pi}\left[K_{\alpha}^{-+}(-\theta)K_{\beta}^{+-}(\theta)+K_{\alpha}^{+-}(-\theta)K_{\beta}^{-+}(\theta)\right]e^{-2ML\cosh\theta}+\dots\]
 where, the boundary fugacities can be expressed in terms of the soliton/anti-soliton
reflection factors: $K_{\alpha}^{+-}(\theta)=R_{+}^{+}(i\frac{\pi}{2}-\theta)_{\alpha}$
and $K_{\alpha}^{-+}(\theta)=R_{-}^{-}(i\frac{\pi}{2}-\theta)_{\alpha}$.

If, however, we are in the attractive domain then the leading finite
size correction is given by the one particle boundary coupling terms
of the breathers:\[
E_{0}(L)=E_{0}(\infty)-m_{n}\frac{g_{\alpha}^{n}g_{\beta}^{n}}{4}e^{-m_{n}L}+\dots\quad;\qquad m_{n}=2M\sin\left(\frac{np\pi}{2}\right)\]
 If the one particle terms of the lightest particle, (the first breather),
are non-vanishing $g_{\alpha}^{1}g_{\beta}^{1}\neq0$ then the corresponding
term provides the leading finite size correction. If any of them is
zero (symmetric boundary with $\phi_{0}=0$) then the leading finite
size correction is determined by the second breather's term since
$g_{\alpha}^{2}$ is never vanishing.

We are going to recover this behavior from the ground-state NLIE separately
for the attractive and in the repulsive regimes. In the repulsive
case the ground state energy can be described either by the pure NLIE
for $H_{+}>0$ or by including the source term corresponding to an
imaginary root for $H_{+}<0$. Since the analysis was already done
in the first case in \cite{BLusch} we focus on the second possibility.
The asymptotic form of the NLIE for large volume can be written as

\[
Z\left(\theta\right)=2ML\sinh\theta+P_{{\rm \mathrm{bdry}}}\left(\theta\right)-\chi(\theta-\theta_{0})-\chi(\theta+\theta_{0})\qquad;\quad e^{iZ(\theta_{0})}=1\]
 where we neglected the exponentially small corrections coming from
the convolution term. We plug this expression into the energy formula
\[
E=-M\cosh\theta_{0}-M\Im m\int_{-\infty}^{\infty}\frac{d\theta}{2\pi}\sinh(x+i\eta)\log(1-e^{iZ(\theta+i\eta)})\]
 and shift the integration contour to $\eta=\frac{\pi}{2}$. In doing
so we need to know the analytic structure of $P_{{\rm bdry}}(\theta)$.
Using the relation coming from the soliton quantization condition
\cite{DNLIE} we can rewrite the boundary fugacity in terms of the
soliton reflection factors on the two boundaries and the soliton-soliton
scattering as in (\ref{PRRS}). This provides the analytic continuation
into the domain where the original integral representation is not
valid. For $H_{+}<0$ the appearing (excited state) reflection factor
has a pole at $i\nu_{0}$. This pole is exponentially closely accompanied
with a logarithm of zero singularity at $\theta_{0}$. In shifting
the contour we take care of these singularities by encircling them
with the contour. We use that $\oint\frac{d\theta}{2\pi}\frac{dg(\theta)}{d\theta}\log(f(\theta))=\pm ig(iu_{\pm})$,
(where $\pm$ applies whenever at $u_{\pm}$ the function $f$ has
a pole/zero) and obtain the contribution of the singular terms: \[
-M\cos\nu_{0}+M\cosh\theta_{0}\]
 The volume ($\theta_{0}$) dependent terms cancel, the term $-M\cos\nu_{0}$
gives contribution to the boundary energy ($E_{0}(\infty)$) while
the integral term with its contour shifted to $i\frac{\pi}{2}$ gives
the same integral term it gave in the $H_{+}>0$ case and reproduces
the expected correction. The cancellation of the volume dependent
terms is the consequence of the fact, that the ground state NLIE with
the source term ($H_{+}<0$) can considered as the analytic continuation
of the ground state (and pure) NLIE from the $H_{+}>0$ domain in
$H_{+}$.

In the attractive regime things are more complicated even for the
$H_{+}>0$ case, where there is now BBS in the spectrum. Even if $e^{iP_{{\rm bdry}}(\theta)}$
does not contain boundary dependent poles in the physical strip it
has poles which correspond to boundary Coleman-Thun mechanisms \cite{BCT}.
Both the reflection factors and the bulk scattering matrix contain
Coleman-Thun type poles at $u_{+}^{n}=i\pi np/2$ corresponding to
on-shell diagrams presented on Figure 9. 

\begin{figure}[h]
\begin{centering}
\includegraphics[height=3cm]{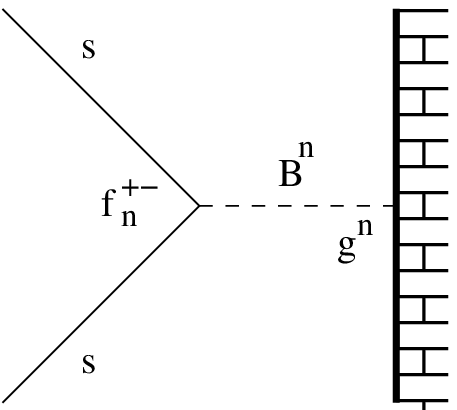}\hspace{3cm}\includegraphics[height=3cm]{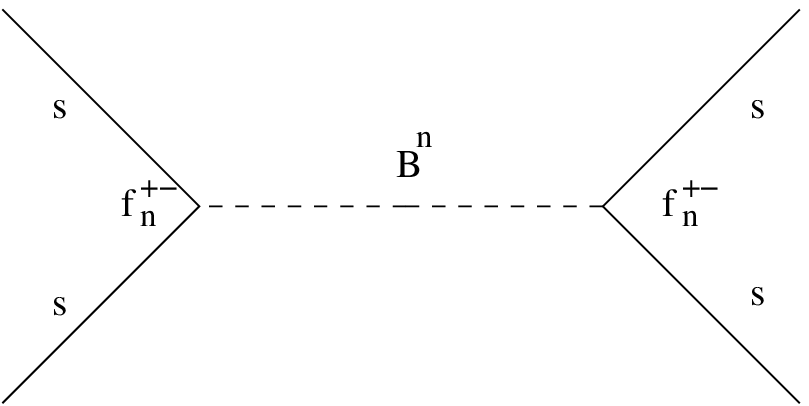}
\par\end{centering}

\caption{Coleman-Thun diagrams for the reflection and scattering matrices}

\end{figure}

\noindent Since the residues are \[
R_{+}^{+}(\theta)_{\alpha}\propto-\frac{i}{2}\frac{f_{n}^{+-}g_{\alpha}^{n}}{\theta-iu_{+}^{n}}\quad;\qquad a(2\theta)\propto-i\frac{f_{n}^{+-}f_{n}^{+-}}{2\theta-2iu_{+}^{n}}\]
 we conclude that $e^{iP_{{\rm bdry}}(\theta)}$ has single poles
at $\theta=iu_{+}^{n}$ with residues \[
e^{iP_{{\rm bdry}}(\theta)}\propto\frac{i}{2}\frac{g_{\alpha}^{n}g_{\beta}^{n}}{\theta-iu_{+}^{n}}\]
 In the exponentially small neighborhood of these poles there are
also logarithm of zero singularities in the energy integral at \[
(1-e^{iP_{{\rm bdry}}(iu_{-}^{n})+2iML\sinh(iu_{-}^{n})})=0\quad;\qquad u_{-}^{n}\approx u_{+}^{n}+\frac{g_{\alpha}^{n}g_{\beta}^{n}}{2}e^{-m_{n}L}\]
 We calculate the contributions from the poles and zeros as before
and obtain the terms \[
-M\sum_{n}\left[\cos(\theta+iu_{+}^{n})-\cos(\theta+iu_{-}^{n})\right]=-\sum_{n}m_{n}\frac{g_{\alpha}^{n}g_{\beta}^{n}}{4}e^{-m_{n}L}\]
 which exactly reproduces the breather's one particle contributions
in the range $H_{+}>0$, since the contribution of the shifted integral
is of order $e^{-2ML}$.

In the $H_{+}<0$ domain we have to include the longest allowed boundary
string to describe the ground state and additionally to take into
account the boundary dependent singularities of $(1-e^{iP_{{\rm bdry}}(\theta)+2iML\sinh(\theta)})$
when we shift the contour. Since the NLIE with the source terms can
be considered as the analytic continuation of the pure NLIE from the
$H_{+}>0$ domain we can see that the terms coming from the accompanying
zeros of the boundary dependent singularities of $e^{iP_{{\rm bdry}}}$
cancel with the volume dependent string energies as it was the case
in the repulsive regime.

\section{UV Behavior}

In this section we compute the ultraviolet (UV) behavior of the various
energy levels.

As $l:=ML\to0$, one can calculate the Casimir energy analytically
by using the asymptotic solution of the NLIE (\ref{NLIE}). From this
it is possible to extract the effective central charge defined by
\[
c_{\mathrm{eff}}(l)=-\frac{24L}{\pi}E(L)\]
 where $E(L)$ is given by Eq.(\ref{theenergy}). In the UV limit,
one can show that only roots and holes growing as $-\log l$ can contribute
to $c_{{\rm \mathrm{eff}}}$. Rescaling the roots and rapidities as
\[
\theta\to\theta-\log l\]
 and introducing the kink counting function $Z_{+}(\theta)=Z(\theta-\log l)$
together with \begin{equation}
Q_{+}(\theta)=2{\rm \Im m}\log\left(1-e^{iZ_{+}(\theta)}\right),\label{defQplus}\end{equation}
 one can express the effective central charge as \[
c_{{\rm \mathrm{eff}}}(0)=\frac{12}{\pi}\left[-\sum_{k}c_{k}e^{\theta_{k}}+\int_{-\infty}^{\infty}\frac{d\theta}{2\pi}e^{\theta}Q_{+}(\theta)\right].\]
 In addition, the NLIE can be rewritten for the kink counting function
$Z_{+}(\theta)$ as \[
Z_{+}(\theta)=e^{\theta}+g_{+}(\theta|\{\theta_{k}\})+\sigma-2i\mathrm{Im}\int dxG(\theta-x-i\epsilon)\log\left[1-e^{iZ_{+}(x+i\epsilon)}\right],\]
 where \[
g_{+}(\theta|\{\theta_{k}\})=\sum_{k}c_{k}\chi_{(k)}(\theta-\theta_{k})\]
 and \[
\sigma=P_{{\rm bdry}}(\infty)+2(2S^{0}+S^{+})\chi(\infty).\]
 Here we defined two integers by \begin{equation}
S^{a}=N_{H}^{a}-2N_{S}^{a}-M_{C}^{a}-2M_{W}^{a}{\rm step}(p-1),\qquad a=0,+,\label{Szeroplus}\end{equation}
 where $N_{H}^{0}$ is the number of holes which do not grow in the
$l\to0$ limit, etc.

By following the standard NLIE method \cite{LMSS}, one can derive
an expression for $\Delta_{0}$, defined by $c_{{\rm \mathrm{eff}}}=1-24\Delta_{0}$.
It is given by \begin{equation}
\Delta_{0}=\frac{1}{8\pi^{2}}\frac{2p}{p+1}\left[P_{{\rm bdry}}(\infty)+\pi+2\pi(K+S^{0})+2\pi\frac{p+1}{2p}S\right]^{2}.\label{Delta0}\end{equation}
 Here the integer $K$ is introduced to relate $Q_{+}(-\infty)$ to
$Z_{+}(-\infty)$ by \[
Q_{+}(-\infty)=Z_{+}(-\infty)+\pi+2\pi K,\]
 using the definition of $Q_{+}$ in Eq.(\ref{defQplus}). Also $S\equiv S^{0}+S^{+}$
can be expressed from Eq.(\ref{Szeroplus}) by \begin{equation}
S=N_{H}-2N_{S}-M_{C}-2M_{W}{\rm step}(p-1).\label{Stotal}\end{equation}
 Since $Q_{+}(-\infty)$ should be given in the fundamental domain
of the $\log$ function, the integer $K$ should be fixed in such
a way that \begin{equation}
-\pi<Q_{+}(-\infty)\le\pi.\label{funddomain}\end{equation}

One can relate (\ref{Delta0}) to that of the $c=1$ conformal field
theory with Dirichlet boundary condition with compactifying radius
$R$ given by \[
R=\frac{\sqrt{4\pi}}{\beta}=\sqrt{\frac{p+1}{2p}}.\]
 which describes the UV limit of the Dirichlet sine-Gordon model.
One can easily calculate the boundary term from Eq.(\ref{Pbdry})
\[
P_{{\rm bdry}}(\infty)=\pi+\pi\frac{2p}{p+1}\left[{\rm sign}(H_{+})+{\rm sign}(H_{-})-\frac{H_{+}+H_{-}+2}{p+1}\right].\]
 Using this and Eq.(\ref{boundparams}), in Eq.(\ref{Delta0}) the
conformal dimension becomes \begin{equation}
\Delta_{0}=\frac{1}{2}\left[\frac{\phi_{+}-\phi_{-}}{\sqrt{\pi}}+mR+\frac{1}{R}(K+S^{0}+1)\right]^{2},\label{confdim}\end{equation}
 where the winding number $m$ is defined by \[
m=\frac{1}{2}({\rm sign}(H_{+})+{\rm sign}(H_{+}))-1-S.\]
 Writing here $S$ in terms of the holes and imaginary roots as in
Eq.(\ref{Stotal}) one can see that this is the continuum counting
equation introduced in Eq.(\ref{conticount}).

For the Dirichlet boundary condition, the momentum mode (i.e. the
term proportional to $1/R$) in the conformal dimension (\ref{confdim})
must vanish. This gives a condition \[
K+S^{0}+1=0\]
 which fixes the integer $K$. If this condition is met, Eq.(\ref{funddomain})
can be written as \begin{equation}
\delta-\frac{3}{2}+\frac{p}{p+1}<S^{0}<\delta-\frac{1}{2}+\frac{p}{p+1},\label{newcond}\end{equation}
 where $\delta$ defined by \[
\delta=\frac{s_{+}+s_{-}}{2}-\frac{\gamma}{2\pi}(H_{+}+H_{-})\]
 takes values in the domain $-1<\delta<1$. With this bound, Eq.(\ref{newcond})
restricts possible values of $S^{0}$ strongly. For the repulsive
case $p>1$, the allowed values are $S^{0}=-1,0,1$ while they are
$S^{0}=-2,-1,0$ for the attractive case $p<1$. As a special example,
let us consider a case where only imaginary roots exist. Since these
roots can not have large real parts in the UV limit, the number of
these roots should be identified with $-S^{0}$. This means a possible
number of imaginary roots in the repulsive case is either 0 or 1,
which is consistent with the IR analysis in sect. 3.

\section{Conclusions}

In this paper we investigated the NLIE involving purely imaginary
roots for the DSG model on a finite interval $L$. We were interested
in describing the DM BBSs thus we investigated the case when the boundary
parameters at one end of $L$ were \lq\lq trivial'' (i.e. excluded
the existence of DM type bound states) but on the other end admitted
such a state. We found an exact match between the set of DM type bound
states and bound state solutions of the NLIE albeit sometimes the
correspondence was surprising: it turned out that in certain parameter
domains the pure NLIE (i.e. the one without imaginary roots) describes
an excited state and one has to add certain appropriate root(s) to
get the ground state. We established the equivalence by studying the
large $L$ solutions of the NLIE from which we extracted not only
the energies but also the reflection factors. In this process we exploited
heavily the fact that sometimes the correct NLIE depends on the second
determination of certain quantities. We confirmed our findings by
calculating the boundary Lüscher corrections for the ground states
and by demonstrating that the UV limit of our NLIE reproduces correctly
the conformal dimensions of the expected $c=1$ BCFT.

With these achievements in hand one can certainly think of the following
problems for future research: first the numerical investigation of
these NLIEs to get information about the finite volume behavior of
the bound states that asymptotically correspond to DM. Second the
extension of the NLIE to the case of two non trivial boundary conditions
at the ends of the interval $L$ may also prove interesting: in this
case one expects that, for large $L$ at least, certain pairs of DM
bound states appear in the spectrum. Recently, using semi-classical
quantization for the DSG model, an interesting restriction (\lq matching
rule') was derived for the allowed pairs in \cite{strip}. The semi-classical
procedure in the theory with the more general perturbed Neumann type
boundary condition revealed the existence of some critical volumes
$L_{{\rm crit}}$ beyond which the bound states ceased to exist. It
would be interesting to see whether these statements are valid in
the full quantum theory, i.e. whether they are valid for the solutions
of the new NLIE. The first step in this direction is to generalize
the present discussion to the case when a constraint is satisfied
between the two boundary conditions allowing a BAE type solution of
the model \cite{Raf3,Cao,RafFran}. The ground-state NLIE in this
case was formulated in \cite{GenNLIE} while the hole excited states
were analyzed in \cite{GenexNLIE}. Thus there is an evident need
for proceeding with the description of the BBS which shows the same
pattern as the Dirichlet one, see \cite{GenBB} for closing the boundary
bootstrap in this case.

The boundary sine-Gordon theory is not the only one exhibiting a complex
pattern of boundary excited states. Its supersymmetric generalization
has an even more complex BBS spectrum \cite{SusyBB} and their description
based on the generalization of the ground-state NLIE derived in \cite{SusyNLIE}
is also an interesting problem.

\subsection*{Acknowledgments}

The authors would like to thank R. Nepomechie and G. Takács for the
illuminating discussions and for taking part in the early stages of
this work. This research was partially supported by the Hungarian
research funds OTKA K60040 and by a cooperation between the Hungarian
Academy of Sciences and the Korean KOSEF. CA was supported in part
by a Korea Research Foundation Grant funded by the Korean government
(MOEHRD) (KRF-2006-312-C00096) and ZB was supported by a Bolyai Scholarship
and the EC network {}``Superstring''. FR thanks partial financial
support from the INFN Grant TO12, from the Italian Ministry of University
and Research through a PRIN fund and from the NATO Collaborative Linkage
Grant PST.CLG.980424.


\begin{thebibliography}{10}
\bibitem{GZ} S. Ghoshal and A. B. Zamolodchikov, \textit{Int. J.
Mod. Phys.} \textbf{A9}, 3841 (Erratum-ibid. \textbf{A9}, 4353) (1994)
{[}\texttt{hep-th/9306002}].

\bibitem{DdV} C. Destri and H. de Vega, \textit{Phys. Rev. Lett.}
\textbf{69}, 2313 (1992) {[}\texttt{hep-th/9203064}]; \\
 C. Destri and H. de Vega, \textit{Nucl. Phys.} \textbf{B438}, 413
(1995) {[}\texttt{hep-th/9407117}].

\bibitem{KBP} A. Klümper, M. T. Batchelor and P. A. Pearce, \textit{J.
Phys.} \textbf{A24}, 3111 (1991).

\bibitem{FMQR} D. Fioravanti, A. Mariottini, E. Quattrini and F.
Ravanini, \textit{Phys. Lett.} \textbf{B390}, 243 (1997) {[}\texttt{hep-th/9608091}]. 

\bibitem{DdV97} C. Destri and H. de Vega, \textit{Nucl. Phys.} \textbf{B504},
621 (1997) {[}\texttt{hep-th/9701107}].

\bibitem{LMSS}A. LeClair, G. Mussardo, H. Saleur and S. Skorik, \textit{Nucl.
Phys.} \textbf{B453}, 581 (1995) {[}\texttt{hep-th/9503227}].

\bibitem{GenNLIE} C. Ahn and R. I. Nepomechie, \textit{Nucl.Phys.}
\textbf{B676}, 637 (2004) {[}\texttt{hep-th/0309261}].

\bibitem{DNLIE} C. Ahn, M. Bellacosa and F. Ravanini, \textit{Phys.
Lett.} \textbf{B595}, 537 (2004) {[}\texttt{hep-th/0312176}].

\bibitem{GenexNLIE} C. Ahn, Z. Bajnok, R. I. Nepomechie, L. Palla
and G. Takács, \textit{Nucl. Phys.} \textbf{B714}, 307 (2005) {[}\texttt{hep-th/0501047}].

\bibitem{MD} P. Mattsson and P. E. Dorey, \textit{J. Phys.} \textbf{A33},
9065 (2000) {[}\texttt{hep-th/0008071}].

\bibitem{Salsko}S. Skorik and H. Saleur, \textit{J.Phys.} \textbf{A28},
6605 (1995) {[}\texttt{hep-th/9502011}].

\bibitem{BLusch} Z. Bajnok, L. Palla and G. Takács, \textit{Nucl.Phys.}
\textbf{B716}, 519 (2005) {[}\texttt{hep-th/0412192}].

\bibitem{Alcaraz}F. C. Alcaraz, M. Barber, M. T. Batchelor, R. J.
Baxter and G. R. W. Quispel, \textit{J. Phys.} \textbf{A20}, 6397
(1987).

\bibitem{Sklyanin}E. K. Sklyanin, \textit{J. Phys.} \textbf{A21},
2375 (1988).

\bibitem{zam-zam} A. B. Zamolodchikov and Al. B. Zamolodchikov, \textit{Ann.
Phys.} \textbf{120}, 253 (1979).

\bibitem{YL}P. E. Dorey, A. Pocklington, R. Tateo and G. M. T. Watts,
\textit{Nucl.Phys.} \textbf{B525}, 641 (1998).

\bibitem{Chaiho}C. Rim, {}``Boundary massive sine-Gordon model at
the free Fermi limit and RG flow of Casimir energy'', {[}\texttt{hep-th/0405162}].

\bibitem{BRZ} Z. Bajnok, C. Rim and Al. B. Zamolodchikov {}``Sinh-Gordon
Boundary TBA and Boundary Liouville Reflection Amplitude'', {[}\texttt{arXiv:0710.4789}].

\bibitem{Ghoshal}S. Ghoshal, \textit{Int.J.Mod.Phys.} \textbf{A9},
4891 (1994).

\bibitem{BCT}Z. Bajnok, G. Böhm and G. Takács, \textit{Nucl.Phys.}
\textbf{B682}, 585 (2004).

\bibitem{strip} Z. Bajnok, L. Palla and G. Takács, \textit{Nucl.Phys.}
\textbf{B702}, 448 (2004).

\bibitem{Raf3}R. I. Nepomechie, \textit{J.Phys.} \textbf{A37}, 433
(2004).

\bibitem{Cao}J. Cao, H.-Q. Lin, K.-J. Shi and Y. Wang, \textit{Nucl.
Phys.} \textbf{B663}, 487 (2003).

\bibitem{RafFran}R. I. Nepomechie and F. Ravanini, \textit{J.Phys.}
\textbf{A36}, 11391 (2003).

\bibitem{GenBB}Z. Bajnok, L. Palla, G. Takács and G.Zs. Tóth, \textit{Nucl.Phys.}
\textbf{B622}, 548 (2002).

\bibitem{SusyBB}Z. Bajnok, L. Palla and G. Takács, \textit{Nucl.Phys.}
\textbf{B644}, 509 (2002).

\bibitem{SusyNLIE} C. Ahn, R. I. Nepomechie and J. Suzuki, \textit{Nucl.
Phys.} \textbf{B767}, 250 (2007) {[}\texttt{hep-th/0611136}].
\end{thebibliography}
\end{document}